\documentstyle[11pt,aaspp4,flushrt]{article} 

\received{1998 June ...}
\accepted{ }

\slugcomment{To appear in {\it The Astronomical Journal}, Vol.\ 116,
November 1998}

\lefthead{SCHWEIZER \& SEITZER}
\righthead{YOUNG GLOBULAR CLUSTERS IN NGC 7252}

\newcommand{\czhel}{\mbox{$cz_{\rm hel}$}}	
\newcommand{\dvlos}{\mbox{$\Delta v$}}		
\newcommand{\etal}{et al.}			
\newcommand{\hbgdav}{\mbox{$\langle{\rm H}\beta\gamma\delta\rangle$}}
\newcommand{\hi}{\ion{H}{1}}			
\newcommand{\hii}{\ion{H}{2}}			
\newcommand{\hst}{{\it HST\/}}			
\newcommand{\kms}{km~s$^{-1}$}			
\newcommand{\kratio}{K/(H$\epsilon$\,+\,H8)}	
\newcommand{\linerat}{\mbox{$F/F({\rm H}\beta)$}} 
\newcommand{\lineratzero}{\mbox{$F_0/F_0({\rm H}\beta)$}}
\newcommand{\msun}{\mbox{${\cal M}_{\odot}$}}	
\newcommand{\mgb}{Mg$\;b$}			
\newcommand{\n}{NGC~}				
\newcommand{\PA}{{\it PA}}			
\newcommand{\reff}{\mbox{$R_{\rm eff}$}}	
\newcommand{\rtidal}{\mbox{$R_{\rm t}$}}	
\newcommand{\tauphot}{\mbox{$\tau_{\rm phot}$}}	
\newcommand{\tauspec}{\mbox{$\tau_{\rm sp}$}}	
\newcommand{\tcross}{\mbox{$t_{\rm cr}$}}	
\newcommand{\vi}{\mbox{$V\!-\!I$}}		
\newcommand{\vizero}{\mbox{$(V\!-\!I)_0$}}	
\newcommand{\zsun}{\mbox{$Z_{\odot}$}}		

\begin{document}

\title{AGES AND METALLICITIES OF YOUNG GLOBULAR CLUSTERS \\
       IN THE MERGER REMNANT NGC 7252\footnote{
Based on observations made at the Cerro Tololo Inter-American
Observatory, National Optical Astronomical Observatories, operated by AURA,
Inc., under a cooperative agreement with the National Science Foundation.}}

\author{FRAN\c COIS SCHWEIZER\altaffilmark{2}}
\affil{Department of Terrestrial Magnetism, Carnegie Institution of
Washington,\\
5241 Broad Branch Road, N.W., Washington, DC 20015-1305; schweizer@dtm.ciw.edu}

\author{PATRICK SEITZER}
\affil{Department of Astronomy, University of Michigan, 830 Dennison Building,\\
Ann Arbor, MI 48109; seitzer@astro.lsa.umich.edu}

%
\altaffiltext{2}{Guest Observer, Michigan--Dartmouth--MIT Observatory.}


\begin{abstract}
\noindent
Ultraviolet-to-visual spectra of eight young star clusters in the merger remnant
and protoelliptical galaxy \n7252, obtained with the Blanco 4-m telescope on
Cerro Tololo, are presented.  These clusters lie at projected distances of
3\,--\,15 kpc from the center and move with a velocity dispersion of $140\pm 35$
\kms\ in the line of sight.  Seven of the clusters show strong Balmer
absorption lines in their spectra [EW(H$\beta$)~= 6\,--\,13 \AA], while the
eighth lies in a giant \hii\ region and shows no detectable absorption
features.  Based on comparisons with model-cluster spectra computed by
Bruzual \& Charlot and Bressan, Chiosi, \& Tantalo, six of the absorption-line
clusters have ages in the narrow range of 400\,--\,600 Myr, indicating that
they formed early on during the recent merger. These clusters, and probably
also the 7$^{\rm th}$ absorption-line cluster, are {\it globular\/} clusters
as judged by their small effective radii and ages corresponding to
$\sim$10$^2$ core crossing times.  The one emission-line object is
$\la$10 Myr old and may be a nascent globular cluster or an OB association.
The mean metallicities measured for three clusters are {\it solar\/} to
within about $\pm$0.15 dex, suggesting that the merger of two likely Sc
galaxies in \n7252 formed a globular-cluster system with a bimodal
metallicity distribution.  Since \n7252 itself shows the characteristics
of a 0.5\,--\,1~Gyr old protoelliptical, its second-generation
solar-metallicity globulars provide direct evidence that giant ellipticals
with bimodal globular-cluster systems can form through major mergers of
gas-rich disk galaxies.  A puzzling property of the observed young globulars
are their high masses of 1\,--\,35\,${\cal M}$($\omega$ Cen) implied by the
luminosities and ages (for an assumed Salpeter IMF).  A spectrum of a
candidate superluminous globular cluster in the elliptical galaxy \n1700,
obtained with the Hiltner telescope at MDM Observatory, shows this object
to be a foreground star.

\end{abstract}


\keywords{galaxies: abundances --- galaxies: formation --- galaxies:
individual (\n7252, \n1700) --- galaxies: interactions --- galaxies:
star clusters}

\section{INTRODUCTION}

Our knowledge of the formation and evolution of globular-cluster systems is
advancing rapidly at present.  Observations with the {\it Hubble Space
Telescope (HST)\/} have led to a slew of discoveries of young star clusters
in merging and starburst galaxies.  Many of these clusters appear to be
globular in nature based on their compactness, high luminosities, and ages
inferred from broad-band colors.  Spectroscopic observations of select
young clusters tend to confirm this view, but are still sparse.  Even though
in many starburst galaxies the mechanism triggering a burst remains unknown,
the process of cluster formation is becoming clearer.  It now appears that
star clusters in general---and globular clusters specifically---may form
{\it preferentially\/} in high-density regions of starbursts (e.g.,
\cite{meur95}), likely from Giant Molecular Clouds squeezed by the
surrounding hot gas (\cite{jog92}; \cite{elme97}).

Galactic mergers offer a special opportunity for learning more about the
cluster-formation process.  Not only do they appear to have been an integral
part of hierarchical galaxy building in the past (e.g., \cite{lars90}), but
they also continue to the present date (see reviews by Kennicutt, Schweizer,
\& Barnes 1998), produce the most vehement known starbursts (\cite{sand96})
and can lead to the wholesale formation of new subsystems of globular clusters
(\cite{schw87}; \cite{ashm92}, 1998; \cite{whit98}).  Observations of the
cluster systems in recent merger {\it remnants\/} such as \n1275
(\cite{carl98}), \n3597 (\cite{holt96}), \n3921 (\cite{schw96b}), and
\n7252 (\cite{mill97}) are especially valuable because in
such remnants the starbursts have largely subsided, dense gas and dust
hindering optical observations have diminished, and most of the freshly
minted clusters have evolved sufficiently to reveal their true nature.
Over the several 10$^8\>$yr a galactic merger takes to complete, loose star
clusters and associations tend to disperse, while gravitationally bound dense
clusters (globular or open depending upon their richness) survive as a
fossil record of the merger's star-formation history.  Hence, determining
the ages and metallicities of these surviving clusters is an important step
in trying to reconstruct that history.

The discovery of bimodal color and metallicity distributions in the
globular-cluster systems of many giant ellipticals, predicted from merger
models of E formation by Ashman \& Zepf (1992), clearly signals the
occurrence of a second event (e.g., \cite{zepf93}; \cite{whit95};
\cite{geis96}).  Opinions differ whether this second event was a major
merger of two gas-rich galaxies or not (see detailed review by \cite{az98}),
and any additional evidence for or against such a merger origin of cluster
bimodality would be valuable.  Since in elliptical galaxies with bimodal
cluster-color distributions the red globulars typically have near-solar
metallicities (Forbes, Brodie, \& Grillmair 1997; Cohen, Blakeslee, \&
Ryzhov 1998; Kissler-Patig \etal\ 1998), a question of great interest is just
how metal-rich the second-generation globular clusters formed during major
disk--disk mergers are.  Preliminary evidence that such clusters may have
approximately solar metallicities stems from relatively strong metal features
observed in the spectra of two young globulars in \n7252 (\cite{ss93};
\cite{frit95}) and of one such cluster in \n1275 (\cite{zepf95b};
\cite{brod98}). However, this evidence is not based on quantitative
estimates, and the need for such estimates persists.

\begin{figure}
\plotfiddle{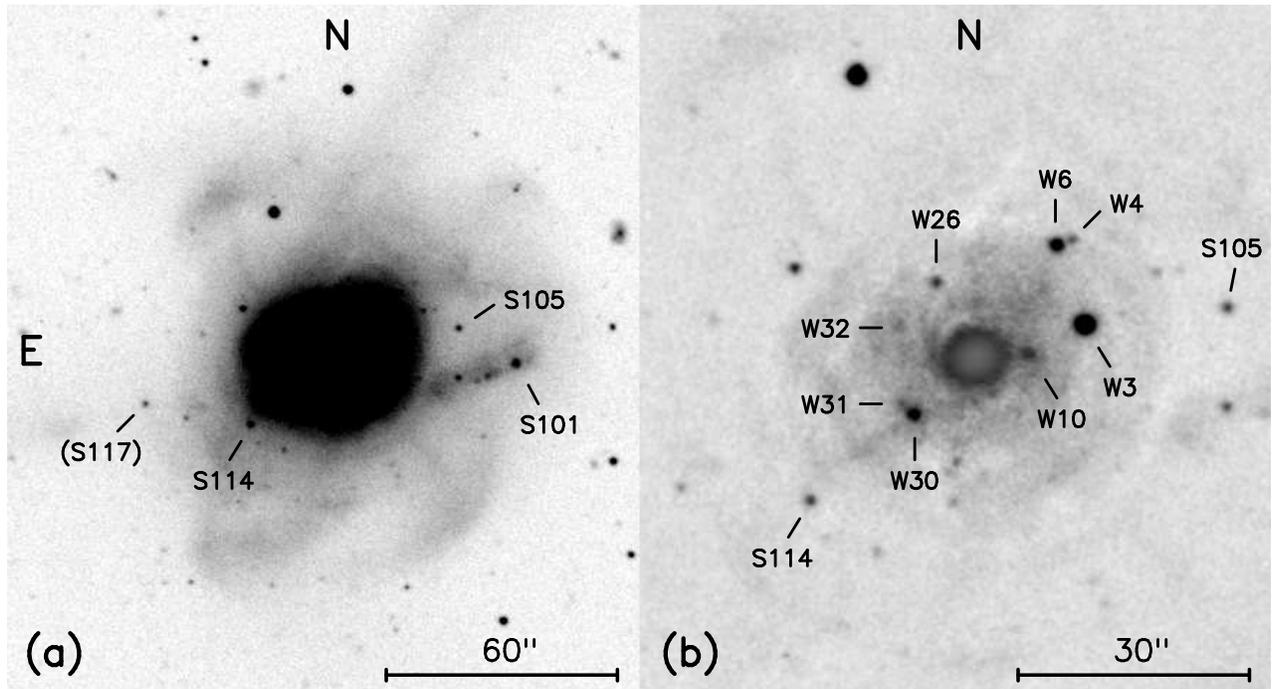}{4.0in}{-90}{74}{74}{-290}{400}
\caption{
Blue-light image of \n7252 with 12 observed candidate clusters marked.
The nine objects yielding useful spectra are listed in Table~\ref{tab1}.
This image was obtained by coadding digital scans of three photographic
plates exposed for a total of 3.2 hr at the prime focus of the Blanco 4-m
telescope. ({\it a}) Direct image, and ({\it b}) masked version, twice
enlarged.
\label{fig101}}
\end{figure}

The present paper describes new spectroscopic observations aimed at deriving
ages and metallicities for a small sample of candidate young globular clusters
in \n7252.  A prototypical remnant of two merged disk galaxies, \n7252
has been observed extensively (see review by \cite{schw98}) and is the only
remnant so far to have been
modeled in detail via $N$-body simulations (\cite{himi95}).  Both the
observations and the model strongly suggest that \n7252 is a 0.5\,--\,1
Gyr old protoelliptical.  Although over the next 10 Gyr the galaxy will
fade by about 1\,--\,1.6 mag from its present-day $M_V = -22.0$, it
is likely to remain a luminous field elliptical.  Young star clusters
of exceptional luminosity were discovered in it first from the ground
(\cite{schw82}; see also Fig.~\ref{fig101} here) and then in larger numbers
with \hst\ (Whitmore \etal\ 1993, hereafter \cite{whit93}).  A recent
\hst\ study of \n7252 yields about 500 candidate clusters more luminous
than $M_V\approx -7.4$ ($V\approx 26.6$), which appear to belong to three
populations: a prominent halo population of young blue globulars, a fainter
halo population of old reddish globulars, and a central disk population
of very young, but more diffuse clusters resembling OB associations
(\cite{mill97}).  Our spectroscopic observations are restricted to candidate
clusters of the young halo population.

\n7252 is located at $\alpha_{\rm J2000} = 22^{\rm h}20^{\rm m}44\fs78$,
$\delta_{{\rm J}2000}=-24\arcdeg40\arcmin41\farcs8$ (\cite{mill97})
and has a recession velocity relative to the Local Group of $+4828$ \kms\
(\cite{schw82}), which places it at a distance of 64.4 Mpc (for $H_0=75$
\kms\ Mpc$^{-1}$, adopted throughout the present paper).  At that distance,
the projected scale is $1\arcsec=312$ pc.  The corresponding true distance
modulus is $(m-M)_0=34.04$, and the apparent visual distance modulus is
$V-M_V = 34.08$ for a Milky Way foreground extinction of $A_V=0.04$
(\cite{rc3}).

In the following, Sec.\ 2 describes the cluster selection and Sec.\ 3 the
observations and reductions.  Section 4 presents results concerning the
dynamics, ages, and metallicities of the observed young clusters, while
Sec.\ 5 discusses various issues concerning the clusters' age distribution,
metallicity, physical nature, and origin.  Finally, Sec.\ 6 summarizes the
main conclusions.

\section{CLUSTER SELECTION}

The candidate clusters to be observed in \n7252 were selected from two
object lists.

The first list is that of 37 ``Outer Clusters'' published by \cite{whit93}.
These candidate globular clusters lie at projected distances of
$5\farcs5 < r < 27\arcsec$ from the center of \n7252 and were situated
within the field of view of \hst's first-generation Planetary Camera
as positioned by Whitmore \etal

The second list contains nine pointlike blue objects lying outside this field
of view at projected distances of $19\arcsec < r < 51\arcsec$ from the center.
These blue objects were among 20 pointlike objects identified on a deep
photographic image of \n7252 obtained with the Blanco 4-m telescope (see
Fig.~\ref{fig101}).  The 20 objects were ranked according to their
apparent color by visually intercomparing the deep image with a map of
the color index $B\!-\!R$\/ produced by Hibbard \etal\ (1994, esp.\ their
Fig.\ 5d) from CCD images.  The nine bluest objects from this sample resemble
in color the candidate young globular clusters identified by \cite{whit93}
with \hst\ and were, therefore, judged to be good candidate globulars
themselves.

A final observing list was prepared by merging the two above object lists,
ranking the objects by their estimated $B$ magnitudes, and selecting all
objects (30) brighter than $B=23$.

\section{OBSERVATIONS AND REDUCTIONS}

Candidate clusters from the above list were observed with the
Ritchey-Chr\'etien spectrograph of the Blanco 4-m telescope on Cerro
Tololo during 1994 September 29 -- October 2.  One and one-half nights were
workable, during which 12 candidate clusters were observed.
Figure~\ref{fig101} identifies these objects.

Of the 12 observed objects, nine yielded useful spectra.  Table~\ref{tab1}
lists these nine objects, of which eight indeed are star clusters in \n7252,
while one is a background galaxy.  Columns 1 and 2 give a running number and
the adopted cluster ID, respectively, Cols.\  3\,--\,6 coordinates relative to
the nucleus of the galaxy, Col.\ 7 the half-light radius \reff\ of each
cluster, and Cols.\ 8 and 9 magnitudes and color indices measured with \hst\
(\cite{mill97}).

For each observation, the slit of the spectrograph was put across at least
two, and occasionally three, candidate clusters at a time.  Suitable cluster
pairs or triplets were selected according to the estimated $B$ magnitudes
and relative position angle.  This angle had to be close to the parallactic
angle at the time of mid-exposure in order to minimize light losses on the
jaws of the $1\farcs6$ wide slit due to differential atmospheric refraction.
The chosen position angles \PA$_{\rm obs}$ and exposure times are given in
Cols.\ 10 and 11 of Table~\ref{tab1}.  To observe any given cluster pair or
triplet,
the spectrograph was first rotated to the required position angle, and the
slit was centered on the galaxy nucleus.  A guide star was then acquired,
and the telescope was offset blindly by moving the guide-star probe by
measured amounts in both coordinates.  This procedure allows precision
offsets to better than $0\farcs2$ accuracy.

Following are brief remarks concerning three observed candidate clusters
that failed to yield usable spectra.
When laid across the two bright clusters W3 and W30, the spectrograph slit
also crossed W10, a relatively bright candidate cluster at $7\farcs3$ from
the nucleus of \n7252.  However, although this cluster was clearly detected
on the two-dimensional spectral image, no useful one-dimensional spectrum
could be extracted because of the bright, spatially variable galaxy background.
Similarly, when crossing the clusters W6 and W26 the spectrograph slit also
crossed candidate Cluster W4 lying $2\farcs1$ west-northwest of W6.  However,
W4 is about 1.8 mag fainter than W6, and its spectrum could not be clearly
disentangled from that of W6.  Therefore, we discuss only the spectrum of
W6, which may contain a small contribution ($\lesssim$10\%) of light from
W4.  Finally, when observing Cluster W31 we positioned the slit to also
cross the much redder candidate cluster W32 ($V\!-\!I = 1.28$ according
to \cite{mill97}).  Due to increasing cirrus during the exposure,
only the first subexposure of 3000 s yielded a well exposed spectral image,
and on this image the spectrum of W32 was too faint in the blue to be of
any use.  The spectrum of W31, on the other hand, is usable though noisy
and yielded a radial velocity.

During our observations, the RC spectrograph was equipped with the KPGL\,\#1
grating and a Reticon~2 chip as a detector.  The slit dimensions were
$1\farcs6\times 314\arcsec$. The recorded spectra cover the approximate
wavelength range $\lambda\lambda$3570--5570, which at the redshift of
\n7252 corresponds to rest wavelengths of $\lambda\lambda$3515--5480.
The spectral resolution ranges from 3.6 \AA\ to 5.1 \AA\ depending on the
wavelength and position along the slit, and the scale along the slit is
$0\farcs924$ per pixel.

The two-dimensional spectral images were processed with IRAF\footnote{
The Image Reduction and Analysis Facility (IRAF) is distributed by the
National Optical Astronomy Observatories (NOAO), which are operated by
the Association of Universities for Research in Astronomy (AURA), Inc.,
under a cooperative agreement with the National Science Foundation.}
through normal bias subtraction and flat-fielding.  To extract a
one-dimensional spectrum for each candidate cluster, the cluster
continuum was traced on each two-dimensional image, and counts from
pixels within $\pm$2.0 pixel from the trace center were coadded. The
combined sky-plus-galaxy background was extracted from pixels 3\,--\,10
pixels on either side of the trace center via a parabolic fit, scaled,
and subtracted from the cluster spectrum.  Following this subtraction, each
individual cluster spectrum was wavelength- and flux-calibrated.  As a last
step, individually calibrated cluster spectra from different images were
summed into one final spectrum for each cluster.

\begin{figure}
\epsscale{0.93}
\plotone{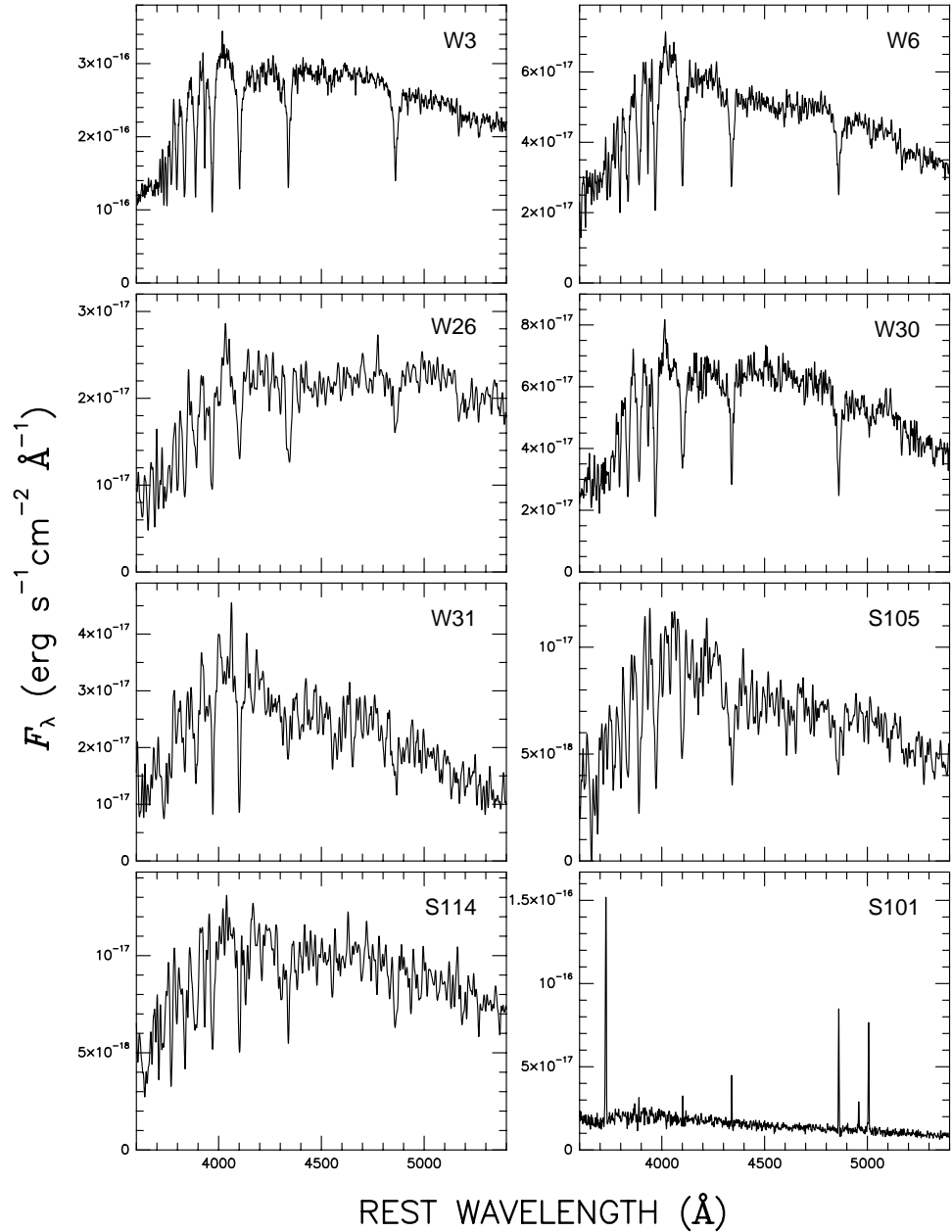}
\caption{
Ultraviolet-to-visual spectra of eight star clusters in \n7252, obtained
with Blanco 4-m telescope plus RC spectrograph through a $1\farcs6$ slit.
The spectra are flux calibrated and plotted versus rest wavelength.
To diminish the appearance of noise, the spectra have been smoothed with
Gaussians of FWHM~= 1, 2, or 4 pixel depending on the signal-to-noise
ratio of their continuum.  Note strong Balmer absorption lines indicative
of A-type main-sequence stars in the first seven clusters, and emission
lines indicative of a very young cluster in the eighth object.
\label{fig102}}
\end{figure}

Figure~\ref{fig102} displays the flux-calibrated sum spectra of the eight
clusters in \n7252, plotted versus rest wavelength. (The ninth object listed
in Table~\ref{tab1}, S117, is a background emission-line galaxy, and its
spectrum is not shown here.)  Note the strong Balmer absorption lines in seven
of the eight clusters, indicative of A-type main-sequence stars.  The eighth
cluster, S101, lies in a giant \hii\ region of the ``Western Loop'' of
\n7252 (\cite{schw82}), and its spectrum is dominated by emission lines.

\begin{figure}
\plotfiddle{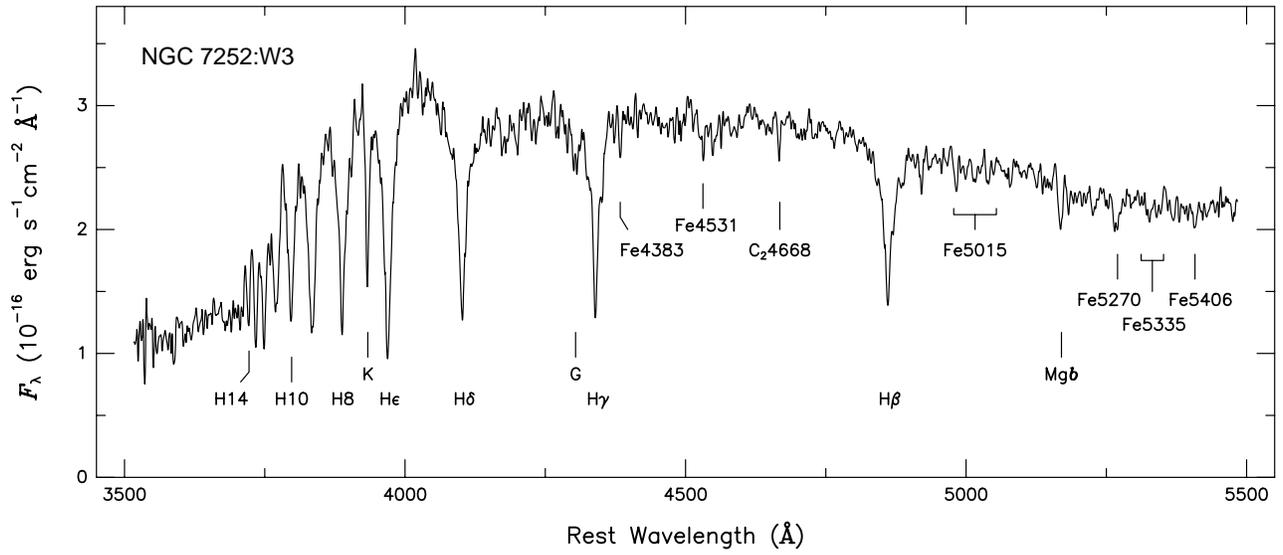}{3.2in}{0}{74}{74}{-290}{-90}
\caption{
Ultraviolet-to-visual spectrum of cluster W3 with main
absorption features identified.  Note the strong Balmer absorption lines up
to H14, the K line of \ion{Ca}{2}, and the relatively strong metal features
surrounding H$\beta$.
\label{fig103}}
\end{figure}

Figure~\ref{fig103} shows the spectrum of the most luminous cluster, W3, on
an expanded wavelength scale and with the main absorption features labeled.
Notice that the Balmer lines H$\beta$ -- H14 are all visible in absorption
and that the metal lines and features, including the Mg triplet centered at
$\lambda$5175, are relatively strong.

\section{RESULTS}

\subsection{Dynamics}

Table~\ref{tab2} gives the heliocentric radial velocities \czhel\ measured for
eight clusters of \n7252.  For the seven clusters with absorption-line spectra,
these velocities were determined by cross-correlating the observed spectra
with a high-signal-to-noise spectrum of the A-type star Kopff~27.  For the
emission-line cluster S101, the velocity was measured directly from the four
emission lines H$\gamma$, H$\beta$, and [O~III]$\lambda\lambda$4959,\,5007;
there are no measurable absorption lines in its recorded spectrum.  The
cluster velocities listed in Col.~2 are mean values formed by averaging the
velocities measured from each subexposure available for any given cluster.
Column~3 gives the standard deviations of the means, and Col.~4 the number
$N$ of subexposures averaged.

As a check, the new velocities for Clusters W3 and W30 can be compared
with velocities measured earlier from spectra obtained with the Palomar 5-m
telescope (\cite{ss93}).  For W3, the new $cz_{\rm hel}= 4821\pm 7$ \kms\
compares well with the previous $4824\pm 23$ \kms.
For W30, the new value of $4624\pm 17$ \kms\ compares with a previous
$4604\pm 31$ \kms. Thus, the radial velocities measured from the new CTIO
absorption-line spectra agree with those measured from the Palomar spectra
to well within the combined $1\sigma$ errors in both cases.

In a similar manner, the newly measured emission-line velocity of
$cz_{\rm hel}= 4504\pm 5$ \kms\ for Cluster S101 compares well with
that of $4509\pm 11$ \kms\ measured for the same \hii\ region by
Schweizer (1982).

To understand the kinematics of any cluster system, the velocities relative
to the host galaxy are needed.  Column 5 of Table~\ref{tab2}
gives the line-of-sight (LOS) velocities \dvlos\ of the clusters relative to
the systemic velocity of \n7252, $cz_{\rm hel,sys} = 4749 \pm 3$ \kms, itself
measured from emission lines of the central ionized-gas disk (\cite{schw82}).  
These relative LOS velocities were computed from
$$\Delta v = (cz_{\rm hel} - cz_{\rm hel,sys})/(1+z_{\rm hel,sys}),$$
where the denominator is a relativistic correction.  The mean relative LOS
velocity of the eight clusters is $\langle\Delta v\rangle = -48 \pm 48$ \kms,
and the LOS velocity dispersion is $\sigma_v = 140 \pm 35$ \kms. These
two quantities have been computed via Pryor \& Meylan's (1993)
maximum-likelihood estimator, and $\sigma_v$ has been corrected for
small-number bias in an approximate manner by applying a multiplicative factor
of $(8/7)^{1/2}$.

The mean LOS velocity of the clusters agrees to within $1\sigma$ with the
systemic velocity of \n7252.  Plots of the individual cluster velocities
versus position show no systematic rotation pattern, whence the velocities
must measure mostly random motions.  The cluster velocity dispersion
of $140\pm 35$ \kms\ is slightly, but not significantly, smaller than the
stellar central velocity dispersion of $\sigma_{\star,0} = 177\pm 12$ \kms\
measured from the calcium triplet at $\lambda$8540 (\cite{lake86}).  For
comparison, note that the $\sigma_v$ of the eight clusters is similar to
that measured for over 80 globular clusters in \n5128 ($\sim${}$140\pm 30$
\kms, \cite{harr88}).

The cluster positions and LOS velocities yield rough estimates of the mass
of \n7252 within the distance to the outermost observed cluster.  We use
the projected mass estimator for a central point mass recommended by
Bahcall \& Tremaine (1981) in the absence of information about the cluster
orbits,
$${\cal M}_0 = \frac{24}{\pi G N} \sum_{i=1}^{N} r_i (\Delta v_i)^2,$$
where $G$ is the gravitational constant, $N$ the number of clusters
(test particles), and $r_i$ the projected distance of the $i$-th cluster
from the center.  With this estimator and the aperture photometry of
Schweizer (1982), the mass is $3.2\times 10^{11}$ \msun\ within
$r\la 15$ kpc (projected distance of most distant cluster, S101), and
the integrated mass-to-visual-light ratio is ${\cal M}/L_V = 7.2$ in
solar units.

If instead of assuming a central point mass, we assumed an extended mass
distribution similar to the number-density distribution of globular clusters,
the estimates would double to ${\cal M}\approx 6\times 10^{11}$ \msun\ and
${\cal M}/L_V\approx 14$ (Heisler, Tremaine, \& Bahcall 1985). However,
the latter estimates are only correct if the tracer population has been
measured either completely or in a representative manner, which is clearly
not the case for the present limited sample of eight globulars (for a
detailed discussion of the uncertainties, see Haller \& Melia 1996).
From checks with subsets of globular clusters we conclude that the
true values of ${\cal M}(r\,${}$\la$\,15 {\rm kpc}) and ${\cal M}/L_V$ are
probably within a factor of two of the first quoted pair of numbers.

Note that the relatively low
${\cal M}/L_V = 7_{-3.5}^{+7} ({\cal M}/L_V)_{\odot}$ within $r\la 15$ kpc
may reflect the post-starburst nature of \n7252's spectrum.  Based on
two-burst models simulating the evolution of two merging Sc galaxies
(\cite{ss92}), \n7252 is predicted to fade by about 1\,--\,1.6 mag
in $V$ over the next 10 Gyr.  Thus, its ${\cal M}/L_V$ should rise to
a value of
${\cal M}/L_V(r\,${}$\la${}$\,4r_{\rm eff})\approx 20$\,--\,30
$({\cal M}/L_V)_{\odot}$ 
more normal for a giant elliptical galaxy (\cite{kent90}).

\subsection{Cluster Ages from Balmer and H\,+\,K Lines}

In principle, the determination of ages and metallicities of single-burst
populations from their spectra is straightforward.  Comparisons between
observed and model spectra, using age- and metallicity-sensitive spectral
features, should yield the answers.  However, there are significant differences
between the model spectra from various groups (see, e.g., Charlot, Worthey,
\& Bressan 1996).
In the present case, the determination of the ages and metallicities of
\n7252 clusters is hampered mainly by the lack of model-cluster spectra of
sufficient spectral resolution for metallicities other than solar.  Whereas
some of the solar-metallicity (hereafter \zsun) models by Bruzual \& Charlot
(1996, hereafter BC96) have the necessary resolution ($\la$2 \AA)
to match the observations, those for non-solar metallicities do not
(wavelength spacings of 10 \AA\ and 20 \AA\ in the UV--optical).  Therefore,
we adopt the following approach.  In the present subsection, we {\it assume\/}
that the clusters have roughly solar metallicity and determine their ages
from the observed spectra via comparisons with the high-resolution model
spectra for $Z=\,$\zsun.  In the next subsection (\S 4.3),
we then estimate metallicities and ages for the best observed two clusters
via traditional methods, finding that---indeed---the assumption of solar
metallicity appears to be good to within a factor of better than two.

To determine cluster ages, we measured the equivalent widths of absorption
lines from the observed spectra and compared them with equivalent widths
measured in exactly the same manner from a set of model spectra.
Table~\ref{tab3} gives equivalent widths of the Balmer and \ion{Ca}{2} H\,+\,K
lines measured from
the observed spectra.  The adopted passbands were 62~\AA\ wide for H$\beta$,
55~\AA\ for H$\gamma$, 52~\AA\ for H$\delta$, 48~\AA\ for H\,+\,H$\epsilon$,
17~\AA\ for K, and 40~\AA\ for H8.  Continuum passbands were chosen on either
side of the line features, and the measurements were carried out with the
task {\it spindex\/} of the image-processing software package VISTA
(\cite{gonz93}).  In addition to the equivalent widths of the measured six
lines, Table~\ref{tab3} also gives the two quantities
\newline\medskip\indent
\hbgdav~$\equiv$ \case{1}{3}[EW(H$\beta$)\,+EW(H$\gamma$)\,+\,EW(H$\delta$)]
\newline
and
\newline\indent
\kratio~$\equiv$ EW(K)/[EW(H\,+\,H$\epsilon$)\,+\,EW(H8)]
\newline\medskip
and the derived logarithmic cluster ages.

Figure~\ref{fig104} illustrates the derivation of cluster ages from the
measured equivalent widths and line ratios.  The plotted curves represent the
evolution of EW(H$\beta$), \hbgdav, and \kratio\ as functions of logarithmic
age, as measured from the model-cluster spectra of solar metallicity
(BC96).  For the Balmer lines, equivalent widths were measured from the
high-spectral-resolution models (``gsHR''), the models based on Gunn \&
Stryker (1983) spectral energy distribution (``gs95''), and the models
based on Kurucz (1995) atmospheres and Lejeune, Cuisinier, \& Buser (1996,
1997) spectra (``kl96''); for details, see BC96.
For the line ratio \kratio, only the high-resolution model spectra
(``gsHR'') could be used because the K line is insufficiently resolved in
the other model spectra.  For comparison with the models, Fig.~\ref{fig104}
also shows the measured equivalent widths and line ratios---marked by 
horizontal lines---of the seven \n7252 clusters with absorption-line spectra.

As Fig.~\ref{fig104} illustrates, the new model-cluster spectra by BC96 now
reproduce quite well the strong Balmer lines observed in several of the
young clusters [EW(H$\beta$)~$\ga$ 10 \AA].  This was not the case with the
older model spectra by Bruzual \& Charlot (1993), as found by Schweizer \&
Seitzer (1993) and Zepf \etal\ (1995b). Note that the strongest observed
Balmer lines are best reproduced by the ``gsHR'' model spectra, presumably
because these spectra have the highest spectral resolution ($\la$2~\AA) among
the three sets of model spectra used.

Cluster ages were determined from Fig.~\ref{fig104} as follows.  Logarithmic
ages were read off at the intersections between the horizontal lines
marking the observed equivalent widths or ratio \kratio\ and the
curves representing the model widths or ratio.  Only measurements with a
significance level of $3\sigma$ or higher were used.  Since for many clusters
the Balmer equivalent widths admit two possible values for the age, the ratio
\kratio\ was used whenever possible to select the more likely of the two
values.  Finally, a weighted mean was formed of the logarithmic ages obtained
from EW(H$\beta$), \hbgdav, and \kratio.  This weighted mean is the logarithmic
age given in the last column of Table~\ref{tab3}.

As Table~\ref{tab3} shows, the ages of six of the eight clusters lie within the
narrow range of about 400\,--\,600 Myr ($\log$\,Age~$\approx 8.6$\,--\,8.8).
The age of Cluster S101, which is still embedded in its \hii\ region, is
$\la$10~Myr, while the relatively poorly observed cluster S114 could be either
about 40~Myr or 1.1~Gyr old.

\subsection{Cluster Metallicities}

To determine cluster metallicities, we measured the Lick line-strength indices
(\cite{fabe85}; \cite{gonz93}) from the observed cluster spectra and
compared them to indices computed for model clusters by Bressan, Chiosi, \&
Tantalo (1996) and BC96.  Our measurements show that only the spectra of
Clusters W3 and W6 have sufficiently high signal-to-noise ratios to
yield Lick indices of the $>${}$3\sigma$ precision required for even coarse
estimates of metallicity.  Hence, the following discussion is restricted to
these two clusters.

The presently available Lick indices for model-cluster spectra of non-solar
metallicity (\cite{bres96}; BC96) are based on analytical fitting functions
derived by Worthey \etal\ (1994).  Because these fitting functions
were designed for stellar populations older than 1 Gyr, one has to be
cautious in using the model indices to interpret spectra of younger clusters.
Therefore, the following discussion proceeds in two steps.  First, we
determine metallicities on the assumption that the logarithmic ages from
Table~\ref{tab3}, themselves derived on the assumption of solar metallicity,
are approximately correct.  This amounts essentially to a consistency check.
Then, in a second step, we make an independent estimate of the cluster
ages and metallicities using the $\log$H$\beta\,$--$\,\log$[MgFe] diagram.

Figure~\ref{fig105} shows six measured line indices of Clusters W3 and W6
superposed on evolutionary tracks for model clusters of five different
metallicities ($Z=0.02$\,--\,2.5 \zsun; \cite{bres96}).  The cluster indices
are plotted at the logarithmic ages given in Table~\ref{tab3}.  Note that
(1) the age-sensitive index H$\beta$ plotted here is the Lick {\it index\/}
and not the equivalent width of Table~\ref{tab3}, and (2) the cluster
indices were measured
from smoothed versions of the observed spectra in order to match the lower
resolution of the Lick indices (for details, see \cite{gonz93}).  The
model evolutionary tracks are plotted only over
the age range over which the fitting functions for $Z=\,$\zsun\ represent
a reasonably good approximation to indices measured directly from a
high-resolution set of model spectra.  As the figure shows, the metallicity
of the two young globular clusters inferred from the five metallicity-sensitive
Fe and Mg indices is roughly solar, to better than a factor of two on
average.

This result is consistent with our finding from Fig.~\ref{fig104} that
the logarithmic ages of W3, W6, and W30 determined from the ratio \kratio\
agree to within 0.1~dex with those determined from the average Balmer
equivalent width \hbgdav\ (taking the higher of the two possible values).
If the cluster metallicity differed by more than a factor of two from solar,
the age- {\it and\/} metal-sensitive ratio \kratio\ would yield ages
significantly different from those derived from Balmer lines.

We have also compared the observed line-strength indices of W3 and W6
with the model-cluster indices of BC96, with very similar results except
in the \mgb\ index.  When interpreted with the BC96 models, this index
would suggest that Clusters W3 and W6 are about twice solar in magnesium
abundance.  This might suggest that, like some giant ellipticals, the
two clusters may perhaps be abnormally rich in magnesium.  However, the
Bressan \etal\ (1996) models show consistency between the Mg and Fe abundances,
whence the Mg enrichment suggested by the BC96 models may simply reflect
model uncertainties.

The $\log$H$\beta\,$--$\,\log[$MgFe] diagram introduced by Gonz\'alez
(1993) yields an independent estimate of cluster ages and
metallicities.  Figure~\ref{fig106} shows this diagram plotted with data
points for W3 and W6 and a grid of isochrones and isofers (lines of equal
metallicity) derived from the Bressan \etal\ (1996) models.  Here, the
quantity [MgFe] is defined through
[MgFe]~$\equiv [{\rm Mg}\;b\times \case{1}{2} ({\rm Fe\,5270} +
{\rm Fe\,5335})]^{1/2}$,
where \mgb, Fe\,5270, and Fe\,5335 are Lick indices expressed in \AA.
This diagram yields cluster ages and logarithmic mean metallicities
relative to the sun of (Age, $[Z]$)~= ($500\pm 20$ Myr, $0.00\pm 0.08$)
for W3 and ($520\pm 30$ Myr, $+0.10^{+0.16}_{-0.19}$) for W6.  Thus, the
cluster ages derived from this diagram agree with those given in
Table~\ref{tab3} to within 10\%, and the relative metallicities $[Z]$ are
indeed close to solar.

Overall, these independent estimates of age and metallicity based on the 
$\log$H$\beta\,$--$\,\log$[MgFe] diagram agree reasonably well with the
cluster ages estimated in \S 4.2 and our former assumption that the clusters
have solar metallicities to within a factor of two.  However, we regard the
ages given in Table~\ref{tab3} (and also Table~\ref{tab5}, see \S5.1) as more
reliable than those derived from the $\log$H$\beta\,$--$\,\log$[MgFe] diagram,
since the latter ages depend on extrapolated fitting functions whereas the
former ages depend on Balmer-line equivalent widths measured directly
from the model spectra.  When $\sim$1~\AA\ resolution model spectra for
clusters of non-solar metallicity become available in the near future,
improved age and metallicity estimates for Clusters W3 and W6 will become
feasible via comparisons with Lick indices also measured directly from the
model spectra.

\subsection{\hii\ Region Around Cluster S101}

Although the strengths of stellar absorption lines in Cluster S101 cannot
be measured from our spectra, the fluxes of the emission lines of the
surrounding \hii\ region (see Fig.~\ref{fig101}) are easily measurable.
Therefore, a chemical abundance can be estimated for this \hii\ region, in
which Cluster S101 formed just recently.  This abundance is presumably
similar to that of the stellar cluster itself and, hence, adds a valuable
third data point for comparison with the abundances of Clusters W3 and W6.

Table~\ref{tab4} gives the observed fluxes $F$ and estimated errors for nine
emission lines ranging from [\ion{O}{2}] $\lambda$3727 to
[\ion{O}{3}] $\lambda$5007.
The errors include internal contributions from continuum fluctuations and
photon noise in the lines themselves, but no external contributions from
uncertainties in the flux calibration.  The table also gives the observed and
reddening-corrected flux ratios relative to H$\beta$, \linerat\ and
\lineratzero, respectively.  The logarithmic reddening constant at H$\beta$
(e.g., \cite{seat79}; \cite{mill96}), $c({\rm H}\beta)\approx 0.64\pm 0.04$,
was estimated from the line ratios $F({\rm H}\gamma)/F({\rm H}\beta)$ and
$F({\rm H}\delta)/F({\rm H}\beta)$ by comparison with a Case A spectrum for
a temperature of $T=10,000$~K (\cite{oste89}) and corresponds to
$E_{B-V}\approx 0.44$ and $A_V\approx 1.3$.

The electron temperature cannot be determined accurately since the
[\ion{O}{3}] $\lambda$4363 line is not detected; however, the upper flux
limit for this line yields an upper limit on the temperature of $T < 15,000$~K
(\cite{oste89}).

To determine the oxygen abundance O/H, we resort to Pagel \etal's (1979)
strong-line method, as developed and enhanced by McGaugh (1991, 1994; see
also \cite{mill96}).  From the line fluxes and ratios of Table~\ref{tab4}
we compute the two quantities
$$\log R_{23} = 0.694\pm 0.035$$
and
$$\log O_{23} = -0.543\pm 0.023,$$
where $R_{23}\equiv$ ([\ion{O}{2}] $\lambda$3727 $+$ [\ion{O}{3}]
$\lambda\lambda$4959, 5007)/H$\beta$ is an empirical indicator of oxygen
abundance and $O_{23}\equiv$ ([\ion{O}{3}] $\lambda\lambda$4959, 5007)/
([\ion{O}{2}] $\lambda$3727) is a measure of the oxygen ionization level.
The above two values, inserted into McGaugh's (1991, esp.\ Figs.\ 9 and 10)
model grids, yield an estimate for the mean ionization
parameter of $\langle U\rangle\approx 5\times 10^{-4}$ and estimates for
the logarithmic oxygen abundance of either
$\log ({\rm O}/{\rm H}) = -3.20\pm 0.05$ or $-4.05\pm 0.06$,
depending on whether the \hii\ region surrounding Cluster S101 falls
on the upper or lower branch, respectively, of the $\log ({\rm O}/{\rm H})$
vs $\log R_{23}$ diagram.  The metallicities corresponding to these two
possible oxygen abundances are approximately $Z = 0.75\pm 0.08$ \zsun\ and
$Z = 0.11\pm 0.02$ \zsun.

To distinguish between these two possible metallicities one needs a
discriminator, for which the flux ratio
$F_0($[\ion{N}{2}] $\lambda$6584$)/F_0($[\ion{O}{2}] $\lambda$3727)
serves well (\cite{mcga94}).  Although our cluster spectra do not cover
the red region of the spectrum and the [\ion{N}{2}] line, two old image-tube
spectrograms of Cluster S101 obtained with the Blanco 4-m telescope
and the same RC spectrograph at dispersions of 25 \AA\ mm$^{-1}$ and
50 \AA\ mm$^{-1}$, respectively, are available (\cite{schw82}).  From
the high-quality image-tube plates, we estimate that
$\slantfrac{1}{5} < F($[\ion{N}{2}] $\lambda$6584$)/F({\rm H}\beta) <
\slantfrac{1}{3}$,
whence the discriminating line ratio is 
$F_0($[\ion{N}{2}] $\lambda$6584$)/F_0($[\ion{O}{2}] $\lambda$3727)$~=
0.19_{-0.04}^{+0.06}$ and its logarithm is $-0.72\pm 0.10$ [using
$F_0({\rm H}\alpha)/F_0({\rm H}\beta) = 2.86$ and the value of
$F_0($[\ion{O}{2}] $\lambda3727)/F_0({\rm H}\beta)$ from Table~\ref{tab4}].
When plotted in the $\log($[\ion{N}{2}]/[\ion{O}{2}]) vs $\log R_{23}$
diagram (\cite{mcga94}, Fig.\ 3), this logarithmic ratio clearly and
unambiguously points toward the \hii\ region of Cluster S101 lying on
the {\it upper\/} branch of the $\log ({\rm O}/{\rm H})$ vs $\log R_{23}$
diagram.  Therefore, the logarithmic oxygen abundance for this \hii\
region, and presumably also for its embedded cluster, is
$\log ({\rm O}/{\rm H}) = -3.20\pm 0.05$, corresponding approximately
to $Z = 0.75\pm 0.08$ \zsun\ or $[Z] = -0.12\pm 0.05$.

The \hii\ region's H$\beta$ luminosity, measured within an area of
$0.5 \times 1.2$ kpc$^2$, is
$L_0({\rm H}\beta) = (7.6\pm 1.0)\times 10^{38}$ erg s$^{-1}$
(Table~\ref{tab4}).
We estimate that the total $L_0({\rm H}\beta)$ within a $1.5\times 1.5$
kpc$^2$ area lies in the range (1\,--\,$2)\times 10^{39}$ erg s$^{-1}$.
Thus, this \hii\ region is once to twice as H$\beta$-luminous as
30~Doradus (\cite{kenn88}) and clearly
ranks among giant \hii\ regions, though not among the most extreme cases
found in spiral galaxies.  For example, it is 2\,--\,4 times as
H$\beta$-luminous as the most luminous \hii\ region in M51, but still
only 1/8$^{\rm th}$ to 1/4 as H$\beta$-luminous as the extreme \hii\ region
in M101 (\cite{sear71}).

\section{DISCUSSION}

The present discussion addresses various issues concerning the ages,
metallicity, and physical nature of the observed young clusters in \n7252.
The two most pressing questions are (1) whether most of these objects are
truly {\it globular\/} clusters and (2) what their relation is to the
metal-rich, but older globular clusters in elliptical galaxies.

\subsection{Cluster Ages}

Given the high luminosities of the three brightest observed globular
clusters ($M_V = -16.2$ for W3, $-14.4$ for W6, and $-14.6$ for W30, if
$H_0 = 75$), the spectroscopically determined cluster ages of \tauspec~=
470\,--\,580 Myr may come as a surprise.  According to the models of cluster
evolution by BC96, a 500 Myr old cluster with a Salpeter IMF fades by only
$\Delta M_V \approx 3.2$ mag over the next 14 Gyr (or, with a Scalo IMF,
by only 2.6 mag).  Hence, even when 14.5 Gyr old these three clusters
should have absolute magnitudes of at least $M_V \approx -11.2$ to $-13.0$.
This would make them more luminous than most presently known old globulars
in the Local Universe.  If we choose $H_0 = 50$, the cluster luminosities
increase by $-0.9$ mag and the problem is aggravated.  Therefore, we need
to carefully check how secure the spectroscopically determined cluster ages
are.

A simple visual comparison of the cluster spectra of Fig.~\ref{fig101} with
a sequence of Magellanic-Cloud cluster spectra arranged by age (Fig.~3 of
\cite{bica86}) suggests immediately that, indeed, the \n7252 clusters have
ages of typically a few 100 Myr.  The spectra of younger Magellanic-Cloud
clusters ($\tau = 10\,$--\,100 Myr) differ significantly from those observed
in the \n7252 clusters by having much stronger UV continua and weaker K lines.
A more detailed comparison with an age sequence of model-cluster spectra
of solar metallicity (BC96) confirms this conclusion.  Whereas the
spectrum of a 570 Myr old model cluster matches the observed spectrum of
W3 extremely well, the spectra of model clusters half or twice
that old show already significant deviations in both line ratios and
continuum shape.  Because of the extreme luminosities of W3, W6, and W30,
we have specifically checked that even during the red-supergiant phase of
cluster evolution ($\tau \approx 10\,$--\,15 Myr) the spectra of model
star clusters remain significantly different (weaker Balmer and \ion{Ca}{2}
K lines) from those observed in the \n7252 clusters.  We conclude that at
least for the most luminous \n7252 clusters the observed spectra allow only
ages of $\tau\ga 300$ Myr. 

As a second check, Table~\ref{tab5} presents a comparison between the
cluster ages determined from our spectra and cluster ages determined
from \hst\ photometry.
Columns~(1) to (3) give the cluster ID, the absolute magnitude $M_V$, and
the color index \vizero\ measured with \hst\ and corrected for foreground
reddening (\cite{mill97}), respectively.  Column~(4) gives the photometric
age \tauphot, derived from \vizero\ via BC96 models and on the assumption
that $Z \approx \,$\zsun.  Finally, Col.~(5) gives the logarithmic age 
determined from the observed spectra (see Table~\ref{tab3}) and Col.~(6) the
spectroscopic age \tauspec\ itself.  Note that for five of the six clusters
with both \tauphot\ and \tauspec, the two ages agree to within the combined
errors.\footnote{
On average, the photometric cluster ages are about 20\% smaller than the
spectroscopic ages.  This difference could be caused by, e.g., a small
systematic error of $\sim$0.03\,--\,0.04 mag in the \vizero\ colors of
the clusters or, conceivably, by non-solar metallicities.  The BC96 models
themselves seem unlikely to cause the discrepancy since the same models
are used to derive both \tauphot\ and \tauspec.}
Thus, the newly derived spectroscopic ages confirm the general cluster
ages estimated earlier from \vi\ colors (\cite{whit93}; \cite{mill97}).
At least for the three \n7252 clusters with the highest signal-to-noise
spectra, we deem the spectroscopic ages to be more reliable than the
photometric ages since they are unaffected by reddening or photometric
zero-point errors.

The spectroscopic cluster ages of Table~\ref{tab5} suggest that many
second-generation globular clusters of \n7252 formed during a relatively
short time interval lasting from about 600 Myr to 400 Myr ago.  Fully six
out of eight age-dated clusters have \tauspec\ lying roughly within this
time interval.  The spectroscopic age of a seventh cluster, S114, is poorly
determined, with formally two possible values of $1100\pm 300$ Myr or
$40_{-24}^{+60}$ Myr.  Within $\pm 2\sigma$ limits this cluster, too, could
have formed during the above time interval.  Finally, the eighth cluster,
S101, was known to lie in a giant \hii\ region of the western loop of
\n7252 before it was observed, whence its very young age ($\la$10 Myr)
comes as no surprise. In this assessment, note that we did not observe
any of the ``inner sample'' clusters within 6\arcsec\ from the nucleus.
Many of these clusters are known to be significantly bluer than the
``outer sample'' clusters (\cite{whit93}) that we did observe and may
be---at least in part---OB associations recently formed within
the central molecular-gas disk (\cite{mill97}).

The above intense cluster-formation period of 600\,--\,400 Myr ago appears
to have occurred in \n7252 {\it shortly after the close encounter\/} of the
two disk galaxies that led to the formation of the present tidal tails and,
eventually, to the galaxies' coalescence.  For $H_0 = 75$ this close encounter
occurred
about 770 Myr ago according to the detailed $N$-body-simulation model of
\n7252 by Hibbard \& Mihos (1995).  Interestingly, when
scaled to \n7252's mass and distance a similar, but generic model of two
merging disk\,$+$\,halo galaxies {\it with gas\/} suggests that the
star-formation rate increased by an order of magnitude over normal beginning
$\sim$100 Myr after the first close approach and lasting for $\sim$140 Myr
(\cite{miho96}).  These times correspond to a period of about 670\,--\,530 Myr
ago.  The close agreement between this {\it predicted\/} period of strongly
enhanced star formation and the 600\,--\,400 Myr period of intense star
formation indicated by the spectroscopic cluster ages of Table~\ref{tab5}
seems remarkable.

\subsection{Metallicity}

The mean metallicities determined for Clusters W3 and W6 and for the
\hii\ region around Cluster S101 are $[Z]\equiv \log(Z/Z_{\odot}) = 0.00\pm
0.08$, $+0.10^{+0.16}_{-0.19}$, and $-0.12\pm 0.04$, respectively (\S\S 4.3
and 4.4).  Thus, all three objects appear to have solar abundances to within
about $\pm$0.15 dex.

These near-solar abundances are perhaps what one might have expected given
that (1) the three clusters formed from molecular gas during the past
0.6 Gyr and (2) the progenitors of \n7252 probably were two luminous,
gas-rich Sc galaxies of $M_V\approx -21$ (\cite{schw82}; \cite{frit94};
\cite{hibb94}).  Yet, measured abundances clearly surpass informed
speculation.

These measured near-solar abundances are also consistent with much recent
evidence that (1) \n7252 itself is a protoelliptical galaxy (e.g.,
\cite{hibb94}; \cite{schw96a}) and (2) giant ellipticals tend to have
globular-cluster systems with bimodal metallicity distributions, one peak of
which contains clusters of roughly solar metallicity (e.g., \cite{whit95};
\cite{zepf95a}; Geisler, Lee, \& Kim 1996; \cite{forb97}).  Observations
with \hst\ indicate that the recent merger
in \n7252 produced several hundred new globular clusters (\cite{mill97}).
The ratio of newly-formed clusters to old clusters is about 0.7, quite close
to the typical value of $\sim$0.5 for the ratio of red (metal-rich) to
blue (metal-poor) globulars in giant ellipticals (Lee, Kim, \& Geisler 1998).
Thus, if most of the newly formed globular clusters in \n7252 are of similar
metallicity as the three above objects, then within a few Gyr \n7252 will
possess a globular-cluster system typical for a giant elliptical. The
metallicity distribution will be bimodal, with a population of metal-poor
clusters consisting of the globulars formerly belonging to the now-merged
disk galaxies and a population of metal-rich clusters formed during the
merger itself.

Note that in \n7252 the metal-poor globulars, identified as objects of
\vi~$\approx$ 1.0 by Miller \etal\ (1997), stem from the merged spiral
galaxies and not from dwarf galaxies accreted by the protoelliptical.
Therefore, at least this one well understood case clearly favors the
Ashman \& Zepf (1992) scenario for the formation of globular-cluster
systems in ellipticals.  It would seem less compatible with the
alternative scenario proposed by Forbes \etal\ (1997) that metal-poor
globulars in ellipticals originated during an early phase of E formation,
unless this early phase explicitly includes the formation of major disk
galaxies that then merge to form the ellipticals themselves.  And it is
quite different from the alternative scenario proposed by C\^ot\'e, Marzke,
\& West (1998), whereby the metal-poor globulars are captured from other
galaxies either through tidal stripping or via the accretion of dwarfs.

\subsection{Nature of Clusters}

What exactly is the nature of the eight star clusters that we have studied
spectroscopically?  Which ones are young {\it globular\/} clusters, and which
ones maybe are not?  And how normal or abnormal are the most luminous of
these clusters which, we claim below, clearly are young globulars?  The
present subsection addresses these questions and some related issues.

The new spectroscopic observations presented above harden the case, made
already by Schweizer \& Seitzer (1993), that many of the bluish point
sources observed in \n7252 by \cite{whit93} and Miller \etal\ (1997)
indeed are young globular clusters.  In essence, any dense cluster with
at least several thousand stars, an effective (= half-light) radius \reff\
of the order of 10~pc or less, and an age exceeding one to two dozen core
crossing times \tcross\ has to be gravitationally bound and is a globular
cluster.

As Fig.~\ref{fig101} and Table~\ref{tab5} show, seven of the eight clusters
studied
feature strong Balmer absorption lines and have ages \tauspec~$\ga 10^8$~yr.  
With such ages and absolute magnitudes of $-12.8 \geq M_V \geq -16.2$,
these clusters must contain well in excess of 10$^5$ stars each (see below).
For five of these clusters (the \cite{whit93} objects), effective radii have
been measured with the refurbished \hst\ and are all
\reff~$< 8$~pc (Table~\ref{tab1}).  Since these five clusters have \reff\ 
typical of Galactic globular clusters, they must have core crossing
times of a few Myr (see, e.g., Meylan \& Heggie 1997) and are, therefore,
typically $\sim$10$^2\,$\tcross\ old (see Table~\ref{tab5}).   Hence, at
least the five objects W3, W6, W26, W30, and W31 are all {\it young globular
clusters\/} beyond any reasonable doubt.\footnote{
Here, as before, we use the term ``young'' for any globular cluster younger
than 1.0~Gyr.}

The two remaining absorption-line clusters, S105 and S114, have been imaged
only from the ground.  On various photographs and CCD frames, their
images are indistinguishable from those of stars.  This puts an estimated
limit of \reff~$\la 0\farcs3$ (95~pc) on their effective radii.  If both
clusters are older than 0.5~Gyr, as their spectra and colors indicate
(Table~\ref{tab5}), then even this weak, ground-based limit puts a rather
stringent upper limit of 0.2 km s$^{-1}$ on any possible systematic expansion
velocity of their stars.  Therefore, even these two clusters are likely
gravitationally bound and bona fide globulars.

Similar arguments cannot be made for the emission-line cluster S101.  Both
its location at the center of a giant \hii\ region and its lack of stellar
absorption lines indicate that this cluster is $\la$10~Myr old.  Although
its high luminosity ($M_V\approx -14.1$), interpreted with various BC96
cluster models, implies a mass in excess of $5\times 10^5$ \msun, its young
age allows no conclusions to be drawn about the long-term stability.  Thus,
this could equally well be a nascent globular cluster or a very massive, but
expanding OB association.

The high luminosities of the absorption-line objects, when combined with the
spectroscopic ages of Table~\ref{tab5} and BC96 models, imply cluster masses
mostly in excess of 10$^7\,$\msun.  This is true for both Salpeter(1955) and
Scalo (1986) stellar initial mass functions (IMF).  Specifically, cluster
models with a Salpeter IMF predict mass-to-light ratios in the range
$0.5 < {\cal M}/L_V < 1.2$ for all objects and individual cluster masses
of $7\times 10^6\,$\msun\ for W31, (1\,--\,4)$\times 10^7\,$\msun\ for W6,
W26, W30, S105, and S114, and an astounding $1.8\times 10^8\,$\msun\ for W3.
Cluster models with Scalo IMF predict masses only about 10\% smaller on
average.  Thus,
all seven absorption-line objects appear currently at least as massive
as $\omega$~Cen ($5\times 10^6\,$\msun; \cite{meyl95}), the most massive
globular cluster of the Milky Way.  Given these extraordinarily high
inferred masses, the question arises whether the most luminous globular
clusters of \n7252 are in any currently detectable way abnormal or not.

Miller \etal\ (1997) note that the five brightest clusters of \n7252
imaged on the PC chip of WFPC2 appear to have more extended wings than
the other clusters.  From the same \hst\ images, we have derived
apparent surface-brightness profiles in $V$ for the three most luminous
clusters W3, W6, and W30 and have compared them with the apparent profiles
of fainter candidate clusters on the same images and of stars measured on
an image of $\omega$~Cen.  When normalized to unity at the center, the
apparent surface-brightness profile of Cluster W3 clearly exceeds the
comparison profiles out to a radius of about 10 pixel (=~$0\farcs46$~=
140 pc), and the profiles of W6 and W30 exceed the comparison profiles out to
$\sim$7 pixel (=~$0\farcs32$~= 100 pc).  For comparison, the median tidal
radius \rtidal\ of Milky-Way globulars is about 35 pc, and only six out of
87 clusters listed by Aguilar \etal\ (1988) have \rtidal~$> 100$ pc.
Grillmair \etal\ (1995) argue that in many cases the tidal radii obtained
by model fitting actually underestimate the true extent of the stellar
distributions in Milky-Way globulars.  Thus, and although we certainly cannot
accurately measure tidal radii for the clusters in \n7252, it would appear
that W3, W6, and W30 are large, but not abnormally large, globulars when
compared to their old counterparts in the Milky Way.

The suggested high masses of ${\cal M}\ga 10^7\,$\msun\ for the most
luminous half dozen globulars in \n7252 need to be checked through
high-resolution spectroscopy and velocity-dispersion measurements
of the kind done by Ho \& Filippenko (1996) for Cluster A in \n1569.
Especially Cluster W3 in \n7252 should be an easy target for 8\,--\,10
meter class telescopes ($V=17.8$).  If the high cluster
masses are confirmed, then the question will be whether such massive
clusters tend to loose 50\%\,--\,90\% of their mass over 15 Gyr due to
mechanisms such as evaporation, tidal stripping, and stellar mass loss
(e.g., \cite{meyl97}).  If they do, then even these highly luminous
clusters might eventually resemble $\omega$~Cen in size, luminosity, and
mass.  If they do not, then extraordinarily luminous and massive
clusters should at least occasionally be found in other external galaxies
and, especially, in ellipticals.

There are very few, if any, extraordinarily luminous {\it old\/} globular
clusters known.  The most luminous, spectroscopically confirmed or
photometrically well measured globular clusters in NGC 1399 and \n4472 have
absolute magnitudes of $M_V\approx -10.9$ (Kissler-Patig \etal\ 1998;
\cite{geis96}).  In M87, at least three likely globular clusters of 
$M_V\approx -11.4$ are known (Whitmore \etal\ 1995).  Scaled relative
to $\omega$~Cen, an old globular cluster of $M_V=-11.0$ should have a
mass of ${\cal M}\approx 1.0\times 10^7\,$\msun.  Therefore, a few old
globulars as massive as the 4$^{\rm th}$\,--\,7$^{\rm th}$ most luminous
young globulars in \n7252 are known, but no old globulars as massive as
the three most luminous globulars in \n7252 are known.

In the dynamically young elliptical galaxy \n1700, Whitmore \etal\ (1997)
point out a bright ($V = 19.6$) point-like source $4\farcs8$ west and
$19\farcs8$ north of the nucleus.  This source has the right color index
(\vi~= 0.93) to be an old globular cluster, but would have an exceptional
$M_V = -13.96$ if it were such a cluster and would be about 3 mag brighter
than the next-brightest candidate globular.  Since the source appears
unresolved, Whitmore \etal\ suggest that instead it may be a star.  On
1996 December 13, we obtained a spectrum of this intriguing object with the
2.4-m Hiltner telescope at the Michigan--Dartmouth--MIT Observatory.  The
4~hr exposure, taken with the Modular Spectrograph in sub-arcsecond seeing
and covering the wavelength range $\lambda\lambda\,$3700\,--\,5600, clearly
shows that this object indeed is a galactic foreground star of spectral
type F or G.

We conclude that either (i) our photometric estimates of masses for the
most luminous young globulars of \n7252 are excessive, or (ii) latter-day
mergers of spiral galaxies occasionally form globular clusters that are
abnormally massive, or (iii) globular clusters very massive at birth
(${\cal M}\ga 10^7\,$\msun) experience significant mass loss over periods of
10\,--\,15 Gyr.  Of these three possibilities, the last seems the most
likely at present (Gnedin \& Ostriker 1997; Portegies Zwart \etal\ 1998).

\subsection{Origin of Young Globular Clusters}

There is growing evidence that young globular clusters in present-day
mergers form from Giant Molecular Clouds (GMC).  Our observations of eight
\n7252 clusters fit in with this evidence, which is as follows.

First, the {\it luminosity\/} functions of young-globular-cluster systems
are remarkably similar to the {\it mass\/} functions of GMCs in present-day
spiral galaxies (\cite{harr94}).  Both kinds of functions are power laws
with similar exponents and upper-end cutoffs.  For example, the luminosity
functions of young clusters in \n4038/39, \n3921, and \n7252 are all of the
form $\phi(L) dL \propto L^{\alpha} dL$, where $-2.1\la \alpha\la -1.6$
(\cite{ws95}; \cite{schw96b}; \cite{mill97}), whereas the mass function of
Galactic GMCs is a similar power law with exponent $-1.63\pm 0.12$.  And
the most massive Galactic GMCs reach masses of about $8\times 10^6$ \msun,
which is only slightly more than the $5\times 10^6$ \msun\ mass of
$\omega$ Cen.

Second, both in \n7252 and in \n3921 the radial distribution of the young
globular clusters follows closely the remnant's visual light distribution,
which itself is well approximated by an $r^{1/4}$-law.  This indicates
that the {\it progenitors\/} of the young globular clusters experienced the
same violent relaxation as did the disk stars (\cite{schw96b}).  This in
turn suggests that these progenitors reacted to the merger like point masses 
and were quite compact.  The GMCs of the input galaxies fit the bill,
while the more diffuse interstellar gas presumably experienced significant
dissipation and piled up near the center of the remnant, redistributing
itself differently from the stars.  This again points to GMCs being the
progenitors of the young globular clusters.

Although not pointing uniquely to GMCs as progenitors, the fact that three
young clusters in \n7252 all have near-solar metallicities (\S 5.2) is
consistent with this picture.  The GMCs of two present-day Sc galaxies of
$M_V\approx -21$ are expected to have about solar metallicities, and nearly
the same must have been true for the two likely Sc galaxies that started
merging in \n7252 about 0.8 Gyr ago (see \S 5.1).  As Jog \& Solomon (1992)
first pointed out, the rapidly mounting pressure of interstellar gas
heated in starbursts provides a natural mechanism for triggering the
collapse of cold GMCs embedded in that gas (see also \cite{elme97}).

It may be no mere coincidence that Cluster S101, the least metal-rich
($Z\approx 0.75$ \zsun) of the three objects measured, is also the youngest
($\la$10 Myr) of the observed clusters and the most distant from the center
($r_{\rm proj}=15$ kpc).  As \hi\ observations and a dynamical model of
\n7252 demonstrate, matter ejected from the two former disk galaxies into
the tidal tails continues falling back into the remnant (\cite{hibb94};
\cite{himi95}).  Matter presently arriving stems from regions farther out
in the input disks than matter that arrived earlier and should, therefore,
be of lower metallicity.  Since Cluster S101 is located in the western
loop, which itself appears to be an inward continuation of the eastern
tail (\cite{schw82}), it may have formed (or just be forming) from a GMC
that was triggered into collapse upon reentry into the denser parts of the
remnant.

Only one result may not quite fit in with the above simple picture of cluster
formation from GMCs.  If confirmed through velocity-dispersion measurements,
the suggested extraordinarily high mass of Cluster W3 exceeds that of any
known Galactic GMC by a factor of about 20.  How any single GMC could have
grown to that mass is unclear, and special processes such as cloud
coalescence or runaway growth in dense shocks may have to be invoked.

With this possible exception, the measured ages and metallicities of young,
mostly globular clusters in \n7252 appear to be consistent with the hypothesis
that these objects formed from GMCs formerly populating the disks of two
late-type, gas-rich galaxies.  Whereas many of these GMCs were triggered into
collapse and cluster formation during the early, most violent phases of the
galactic merger 600\,--\,400 Myr ago, a few late returners may still be
experiencing the same fate at the present time.

\section{SUMMARY}

We have described spectroscopic observations of 12 candidate clusters in
the merger remnant \n7252, nine of which yielded useful spectra.  Of these
nine objects one turns out to be a background galaxy, while the other eight
are high-luminosity star clusters clearly associated with \n7252 itself.
The main results of our analysis are as follows:

(1) The eight clusters lie at projected distances of
$r = 10\farcs4$\,--\,48\arcsec\ (3.2\,--\,15 kpc) from the center and
have a line-of-sight velocity dispersion of $\sigma_v = 140\pm 35$ \kms,
comparable to that of the globular clusters in \n5128 and indicative of
a mass of about $(3.2_{-1.6}^{+3.2})\times 10^{11}$ \msun\ within
$r\la 15$ kpc.  The mass-to-light ratio of \n7252 within the same radius
is ${\cal M}/L_V = 7_{-3.5}^{+7}$ in solar units.

\begin{figure}
\epsscale{0.83}
\plotone{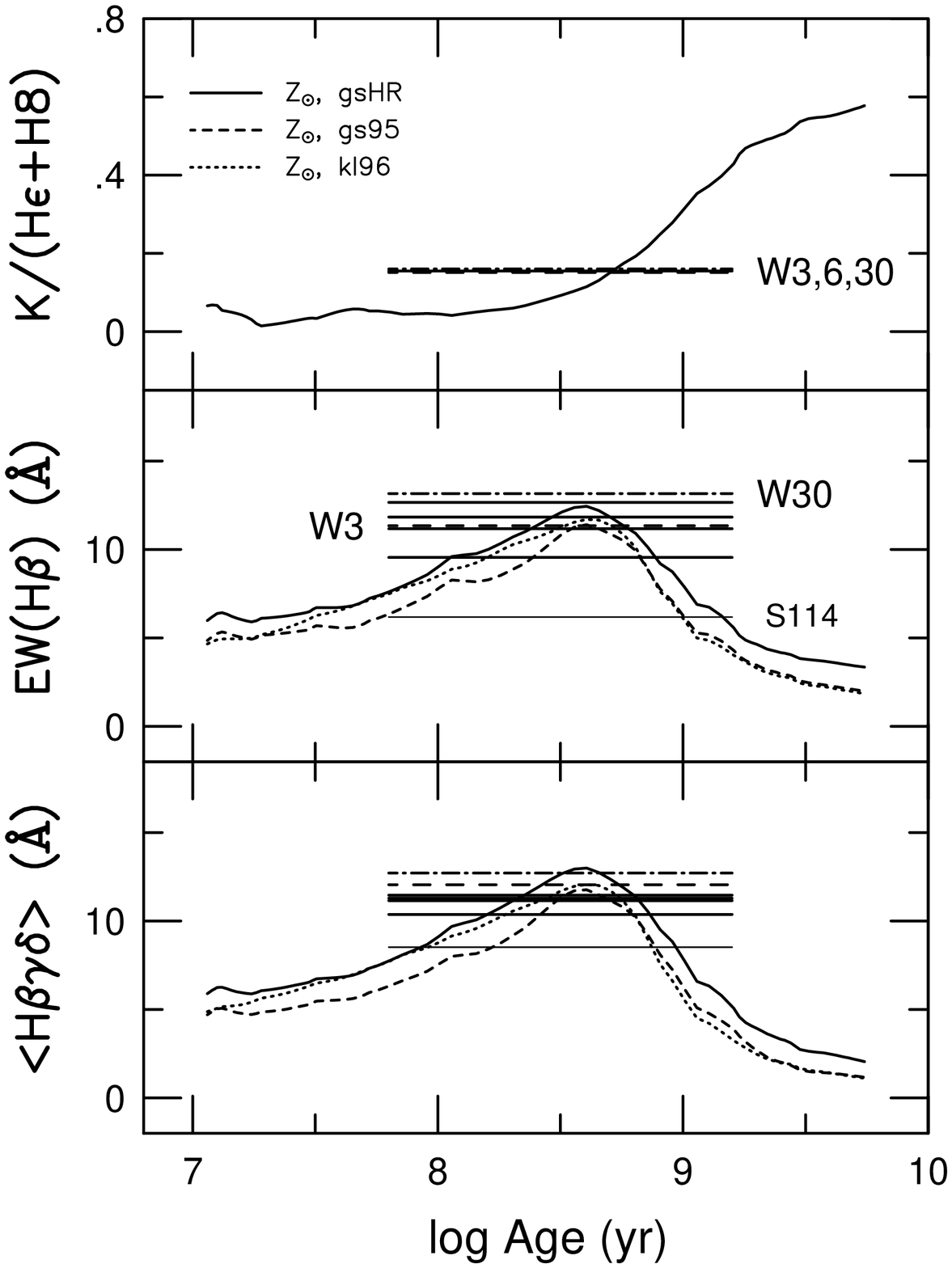}
\caption{
Evolution of line ratio \kratio\ and equivalent widths EW(H$\beta$) and
\hbgdav\ in model-cluster spectra with age ({\it solid, dashed, and dotted
curves}, computed from BC96 models of solar metallicity), compared with
values measured from spectra of seven clusters in \n7252 ({\it horizontal
lines}).  Lines for clusters W3 ({\it dashed\/}), W6, W30 ({\it dash--dot\/}),
and S114 ({\it thin\/}) are marked.  Note that nearly all clusters have ages
falling into the range $10^8$\,--\,$10^9$ yr.
\label{fig104}}
\end{figure}

\begin{figure}
\epsscale{1.0}
\plotone{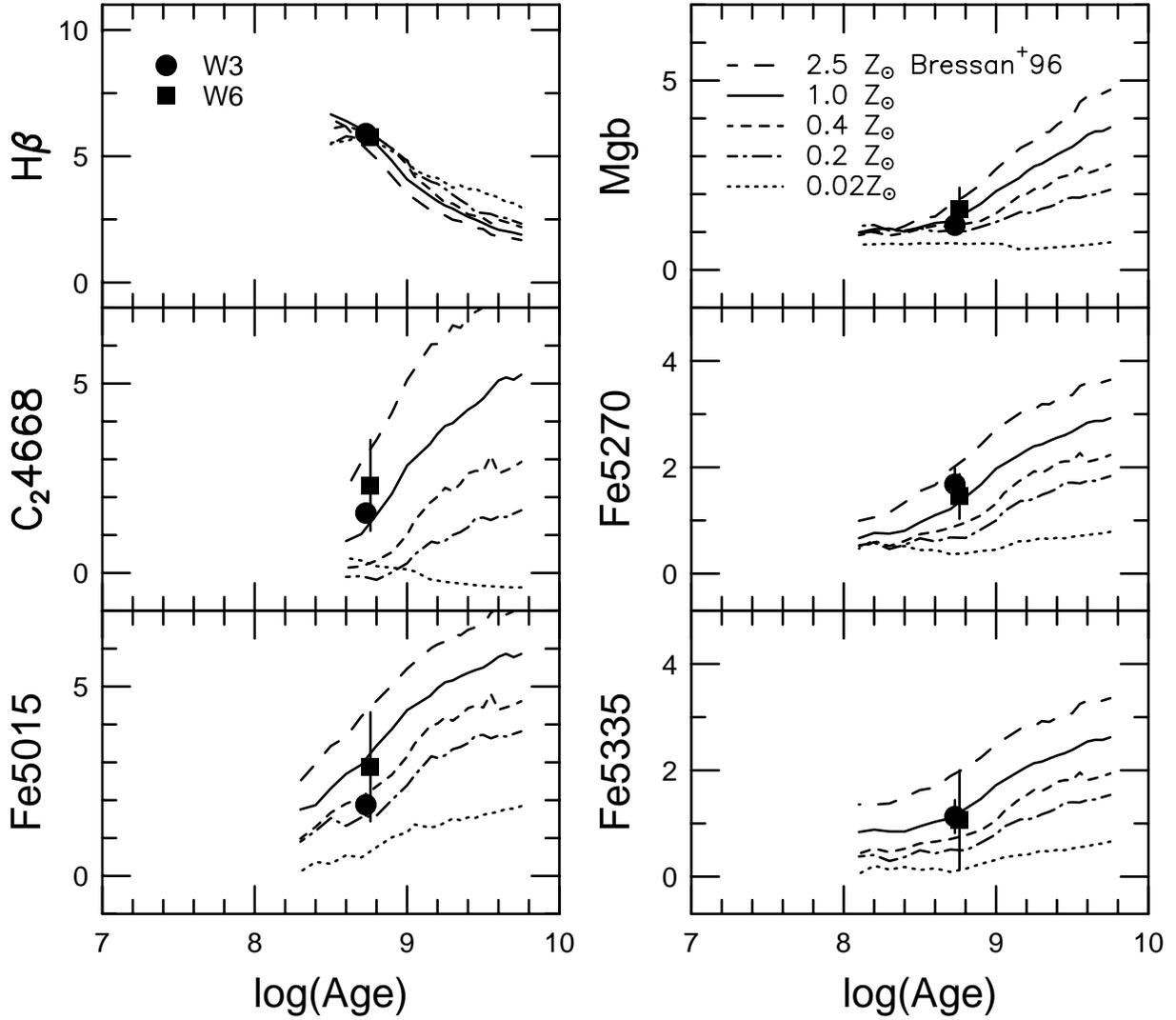}
\caption{
Spectral indices (in \AA) on the Lick system measured for two globular
clusters in \n7252 and compared with spectral-evolution models by Bressan,
Chiosi, \& Tantalo (1996) for five different metallicities ($Z=0.02$\,--\,2.5
\zsun). The data points for W3 and W6 are plotted at the logarithmic ages
given in Table~\ref{tab3} and indicate that the metallicity of these clusters is
roughly solar.
\label{fig105}}
\end{figure}

\begin{figure}
\plotfiddle{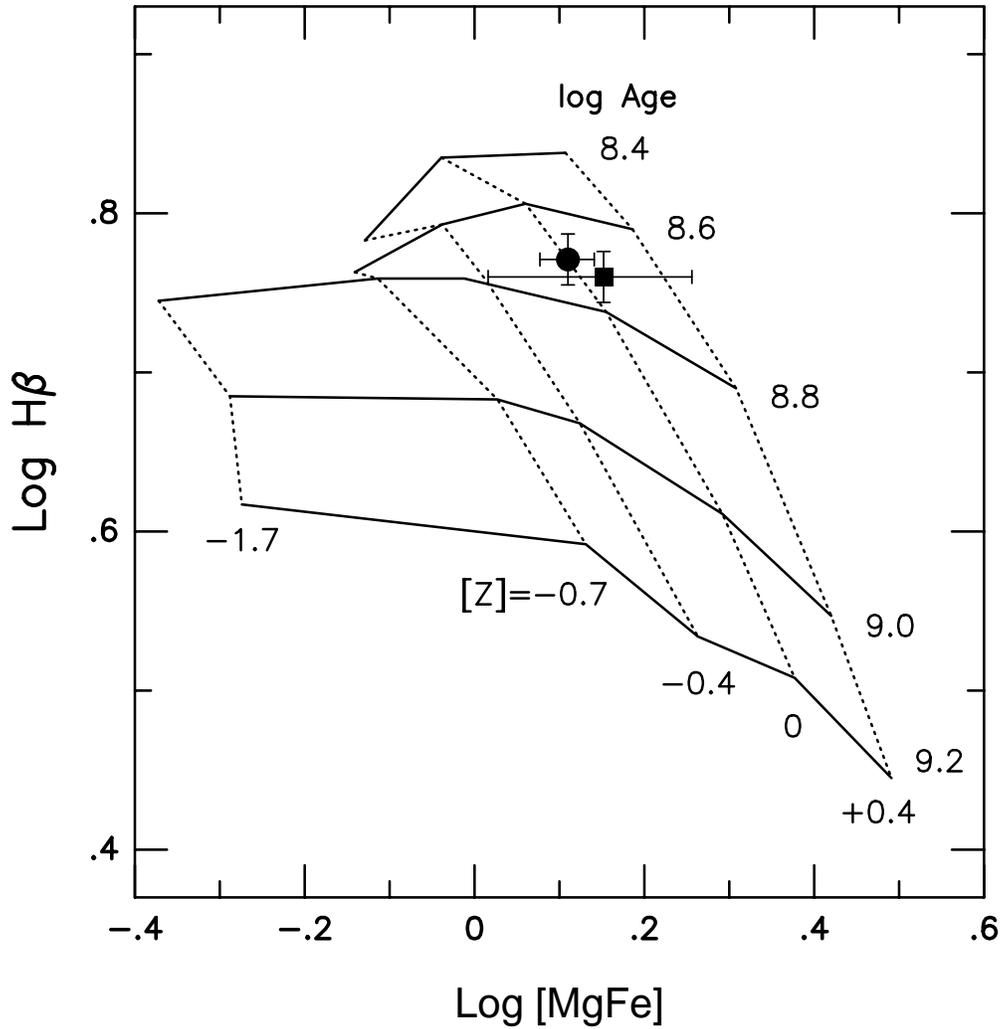}{6.0in}{0}{80}{80}{-240}{-120}
\caption{
Log$\,$H$\beta$ vs $\log$[MgFe] diagram for clusters W3 ({\it filled circle
with error bars\/}) and W6 ({\it square\/}). Grid of isochrones ({\it solid
lines\/}) and isofers ({\it dotted lines\/}) based on Bressan \etal\ (1996)
models is superposed.  From this diagram, the cluster ages appear to be
$\sim$500 Myr and the metallicities close to solar.
\label{fig106}}
\end{figure}

(2) Seven of the eight clusters show strong Balmer absorption lines in
their spectra (H$\beta$\,--\,H14 in the best observed case), while
Cluster S101 lies in a giant \hii\ region and shows no detectable
absorption features.  
In contrast to earlier models, new cluster models by Bruzual \& Charlot
(1996) and Bressan \etal\ (1996) reproduce the observed strong Balmer
lines well (Fig.~\ref{fig104}).  Based on these models, the ages of six of
the absorption-line clusters lie in the narrow range of 400\,--\,600 Myr,
while the age of the seventh cluster is poorly determined.
The emission-line cluster S101 has an estimated age of $\la$10 Myr.

(3) At least five, and probably all seven, absorption-line objects are
young {\it globular clusters}, as judged by their effective radii ($<$8~pc
where measured with \hst) and the fact that they are typically
$\sim$10$^2$ core crossing times old.  The emission-line cluster, being only
a few crossing times old, could be either a nascent globular cluster or a
massive expanding OB association.

(4) The spectra of absorption-line clusters with good signal-to-noise ratios
also feature relatively strong metal lines, including the Mg triplet at
$\lambda$5175.  For two of these clusters and for the gas surrounding the
emission-line object, metallicities can be derived from the spectra.
The mean metallicities are solar to within about $\pm$0.15 dex.  Specifically,
the logarithmic mean metallicities are $[Z] = 0.00\pm 0.08$ for Cluster W3
and $+0.10^{+0.16}_{-0.19}$ for W6, while the oxygen abundance of the
\hii\ region containing Cluster S101 is $\log({\rm O}/{\rm H}) = -3.20\pm
0.05$, corresponding approximately to $[Z] = -0.12\pm 0.05$.

(5) If most of the newly-formed globular clusters in \n7252 have approximately
solar metallicities as Clusters W3, W6, and S101 do, then this recent merger
remnant and protoelliptical has just formed a globular-cluster system with
a bimodal metallicity distribution.  The ratio between its numbers of young
and old globular clusters is about 0.7 (\cite{mill97}), similar to the typical
ratio of $\sim$0.5 between metal-rich and metal-poor globulars observed in
giant ellipticals. Therefore, \n7252 and its globular-cluster system provide
valuable direct evidence for the hypothesis that many giant ellipticals,
and especially those with bimodal cluster systems, formed through major
mergers of gas-rich disk galaxies.

(6) The intense cluster-formation period of 600\,--\,400 Myr ago indicated
by the spectroscopically determined cluster ages seems to have occurred
in \n7252 shortly after the close encounter of two gas-rich disk galaxies
that led to a merger and the present-day remnant.  According to
the dynamical model by Hibbard \& Mihos (1995), this close encounter
occurred about 770 Myr ago ($H_0=75$). It seems likely that the progenitors
of the newly formed, second-generation globular clusters were mostly Giant
Molecular Clouds in the disk galaxies, triggered into collapse and efficient
star formation by surrounding starburst-heated gas.

(7) A puzzling property of the observed young solar-metallicity globulars
are their high masses implied by the luminosities and ages.  When interpreted
with BC96 models featuring Salpeter and Scalo IMFs, the luminosities yield
masses of 1\,--\,8$\times$ that of $\omega$~Cen for six of the globulars
and a whopping $\sim$35$\,{\cal M}$($\omega$ Cen) for the seventh,
exceptionally luminous cluster W3.  For comparison, the most luminous
{\it old\/} globular clusters known in giant ellipticals have estimated masses
of only about 3\,--\,4$\,{\cal M}$($\omega$ Cen).  Thus, either the
inferred masses of the young globulars in \n7252 are too high (e.g., because
of different stellar IMFs) or massive globular clusters experience up to
$\sim$90\% mass loss while evolving from 0.5 Gyr to 15 Gyr.  The predicted
masses of these young globulars can and should be checked through
velocity-dispersion measurements with 8\,--\,10 meter class telescopes.

%


\acknowledgments

We thank Dr.\ M.\ G.\ Smith, Director of the Cerro Tololo Inter-American
Observatory, for allocating us telescope time, Jack Baldwin and German
Schumacher for valuable advice concerning precision offsets, and Hernan Tirado
and the mountain staff for expert technical assistance at the telescope.  
John Hibbard kindly provided a map of the color index $B\!-\!R$\/ in \n7252,
Tad Pryor a copy of his maximum-likelihood-estimator code, and
Alessandro Bressan and Gustavo Bruzual electronic data from their latest
cluster-evolution models.  One of us (F.S.) thanks Sandra Keiser and
Michael Acierno for cheerful and dedicated computer support and gratefully
acknowledges partial support from NSF through Grants AST--92\,21423 and
AST--95\,29263.

%
%
%

\clearpage


\begin{deluxetable}{clccccccccrl}
\def\psn{\phs\phn}
\def\pnn{\phn\phn}
\tablenum{1}
\tablecolumns{12}
\tablewidth{0pt}
\tablecaption{Candidate Clusters Observed in NGC 7252}
\tablehead{
\colhead{}     & \colhead{Other} & \colhead{$\Delta\alpha_{2000}$} &
\colhead{$\Delta\delta_{2000}$}  & \colhead{$r$}   & \colhead{\PA}  &
\colhead{\reff\tablenotemark{b}} & \colhead{$V_{\rm c}$\tablenotemark{b}} &
\colhead{$(V\!-\!I)_{\rm c}$\tablenotemark{b}} & \colhead{\PA$_{\rm obs}$} &
\colhead{Exp.\tablenotemark{c}}  & \colhead{}                      \\
\colhead{\#}   & \colhead{ID\tablenotemark{a}} & \colhead{(\arcsec)} &
\colhead{(\arcsec)} & \colhead{(\arcsec)} & \colhead{(\arcdeg)} &
\colhead{(pc)}      & \colhead{(mag)}     & \colhead{(mag)}     &
\colhead{(\arcdeg)}  & \colhead{(s)}       & \colhead{Type}         }
\startdata
1& W3  & $-$14.56   & \psn3.80   & 15.05&  284.6& $\la$7.1&  17.84&   0.64&  117.5&  7200& GC \nl
2& W6  & $-$11.01   &\phs14.07   & 17.87&  322.0& $\la$4.2&  19.64&   0.64&  107.3&  9600& GC \nl
3& W26 & \psn4.52   & \psn9.37   & 10.40&\phn25.8&$\la$7.6&  20.39&   0.68&  107.3&  9600& GC \nl
4& W30 & \psn7.53   &\phn$-$7.69 & 10.77&  135.6& $\la$5.0&  19.46&   0.63&  117.5&  7200& GC \nl
5& W31 & \psn9.08   &\phn$-$6.29 & 11.05&  124.7& $\la$3.1&  21.07&   0.54&\pnn1.8&  3000& GC \nl
6& S101& $-$48.0\phn&  $-$3.1    & 48.10&  266.3&  \nodata&\nodata&\nodata&\phn58.5&12000& Cl\,+\,H$\>$II \nl
7& S105& $-$33.15   & \psn6.0\phn& 33.69&  280.3&  \nodata&\nodata&\nodata&\phn58.5&12000& GC \nl
8& S114\tablenotemark{d}&
        \phs21.15   & $-$18.7\phn& 28.23&  131.5&  \nodata&  21.23&   0.62&\phn78.9& 9600& GC \nl
9& S117&\phs48.6\phn& $-$13.3\phn& 50.39&  105.3&  \nodata&\nodata&\nodata&\phn78.9& 9600&
                                                                       Galaxy\tablenotemark{e}\nl
\enddata
\tablenotetext{a}{Letter ``W'' identifies objects from Whitmore \etal\ (1993), letter ``S''
	from present paper.}
\tablenotetext{b}{From Miller \etal\ (1997); $V_{\rm c}$ and $(V\!-\!I)_{\rm c}$ are
	corrected for Milky Way foreground extinction of $A_V=0.04$.}
\tablenotetext{c}{Total exposure time of analyzed spectrum, excluding exposures
	discarded because of heavy cirrus.}
\tablenotetext{d}{Cluster S114 is also \#42 in Table 1 of Miller et al.\ (1997).}
\tablenotetext{e}{Background galaxy at redshift $z=0.1091$.}
\label{tab1}
\end{deluxetable}
\clearpage


\begin{deluxetable}{lcccc}
\def\psn{\phs\phn}
\def\pnn{\phn\phn}
\tablenum{2}
\tablecolumns{5}
\tablewidth{0pt}
\tablecaption{Cluster Radial Velocities}
\tablehead{
\colhead{} & \colhead{\czhel} & \colhead{$\sigma_{cz}$} &
\colhead{} & \colhead{\dvlos\tablenotemark{a}}            \\
\colhead{Cluster} & \colhead{(\kms)} & \colhead{(\kms)} &
\colhead{$N$} & \colhead{(\kms)}                           }
\startdata
W3     &  4821 & \phn7 &  2 & \phn$+$71 \nl
W6     &  4709 & \phn9 &  3 & \phn$-$39 \nl
W26    &  4876 &    31 &  3 &    $+$125 \nl
W30    &  4624 &    17 &  2 &    $-$123 \nl
W31    &  4523 &\nodata&  1 &    $-$222 \nl
S101\tablenotemark{b} &
          4504 & \phn5 &  4 &    $-$241 \nl
S105   &  4648 &    37 &  4 & \phn$-$99 \nl
S114   &  4880 &    47 &  3 &    $+$129 \nl
\enddata
\tablenotetext{a} {\dvlos\,=\,(\czhel$-4749)/1.01584$, see text.}
\tablenotetext{b} {Cluster embedded in \hii\ region, em.-line velocity.}
\label{tab2}
\end{deluxetable}
\clearpage


\begin{deluxetable}{lrrrrrrrrc}
\footnotesize
\def\pnn{\phn\phn}
\def\psn{\phs\phn}
\def\pp{$\phantom{\pm}$}
\def\dts{....$\;$\pp ...$\:$}
\def\DTS{....$\;$\pp\pp ....$\phm{a}$}
\tablenum{3}
\tablecolumns{10}
\tablewidth{0pt}
\tablecaption{Line Equivalent Widths and Cluster Ages}
\tablehead{
\colhead{} &                 \colhead{\psn H$\beta$}  & 
\colhead{\phn H$\gamma$} &   \colhead{\phn H$\delta$} &  
\colhead{H\,+\,H$\epsilon$}& \colhead{K} & 
\colhead{H8} &               \colhead{\hbgdav} &
\colhead{} &                 \colhead{} \\
\colhead{Cluster} &          \colhead{\psn (\AA)} &
\colhead{\phn(\AA)} &        \colhead{\phn(\AA)} &
\colhead{\phn(\AA)} &        \colhead{(\AA)} & 
\colhead{(\AA)} &            \colhead{(\AA)} &
\colhead{K/(H$\epsilon$\,+\,H8)}& \colhead{$\log$\,Age\tablenotemark{a}} }
\startdata
W3  & 11.3$\pm$0.3& 11.0$\pm$0.1& 13.8$\pm$0.5& 13.0$\pm$0.1&
	3.5$\pm$0.1& 9.3$\pm$0.5& 12.0$\pm$0.2& 0.156$\pm$0.005& 8.73$\pm$0.02 \nl
W6  &  11.2\pp0.6&   9.5\pp0.6&   12.8\pp0.3&   12.7\pp0.6& 
	3.5\pp1.0&   9.8\pp1.5&   11.2\pp0.3&   0.155\pp0.046&     8.76\pp0.04 \nl
W26 &   9.6\pp2.9&  11.7\pp2.2&   12.6\pp1.5&   11.3\pp1.4&  
	2.3\pp1.1&   9.9\pp1.9&   11.3\pp1.3&   0.110\pp0.054&     8.72\pp0.20 \nl
W30 &  13.2\pp1.0&  11.2\pp1.2&   13.8\pp0.3&   15.5\pp0.1&  
	4.3\pp1.1&  12.2\pp1.3&   12.7\pp0.5&   0.156\pp0.040&     8.67\pp0.04 \nl
W31 &  12.7\pp3.5&   7.9\pp3.8&   13.8\pp5.8&   13.8\pp4.3&  
	\dts     &  11.6\pp2.4&   11.5\pp3.2&           \DTS &     8.58\pp0.23 \nl
S101&$-$18.3\pp2.5&$-$4.5\pp1.5&$-$0.7\pp1.6&        \dts &  
	\dts     &       \dts &        \dts &           \DTS &  $\la$7.0    \nl
S105&  11.8\pp2.0&  13.4\pp3.0&    5.9\pp2.7&   12.8\pp1.0&  
	\dts     &  13.5\pp5.8&   10.4\pp1.5&           \DTS &   8.78\pp0.08\tablenotemark{b} \nl
S114&   6.2\pp1.1&   8.2\pp1.6&   11.2\pp5.4&    9.7\pp3.3&  
	\dts     &   9.5\pp0.7&    8.5\pp1.9&           \DTS &   9.06\pp0.10\tablenotemark{c} \nl
\enddata
\tablenotetext{a}{Age expressed in years.}
\tablenotetext{b}{One of two possible values; other value is $8.36\pm 0.20$.}
\tablenotetext{c}{One of two possible values; other value is $7.6\pm 0.4$.}
\label{tab3}
\end{deluxetable}
\clearpage


\begin{deluxetable}{lcccc}
\def\pnn{\phn\phn}
\def\psn{\phs\phn}
\def\pp{$\phantom{\,\pm\,}$}
\def\ppp{$\phantom{\pm}$}
\def\ple{$\phantom{<}$}
\tablenum{4}				
\tablecolumns{5}
\tablewidth{0pt}
\tablecaption{Emission-Line Fluxes in S101}
\tablehead{
\colhead{}&     \colhead{$\lambda$}&   \colhead{Flux $F$}&  &  \\
\colhead{Line/Ion}&   \colhead{(\AA )}&
\colhead{($10^{-16}$ erg cm$^{-2}$ s$^{-1}$)} &
\colhead{\linerat }&  \colhead{\lineratzero} }
\startdata
[\ion{O}{2}]&     3727&  9.23$\,\pm\,$0.20&  2.62$\,\pm\,$0.22&  3.84$\,\pm\,$0.32 \nl
[\ion{Ne}{3}]&    3869&  0.50\pp 0.25&       0.14\pp 0.07&       0.20\pp 0.10 \nl
H8 $+$ \ion{He}{1}&3889& 0.48\pp 0.25&       0.14\pp 0.07&       0.20\pp 0.10 \nl
H$\delta$&        4101&  0.69\pp 0.24&       0.20\pp 0.07&       0.26\pp 0.09 \nl
H$\gamma$&        4340&  1.39\pp 0.26&       0.40\pp 0.08&       0.48\pp 0.10 \nl
[\ion{O}{3}]&     4363& $<$0.05\pp ......\ple\,&
			               $<$0.014\ppp .....\ple\,&
						         $<$0.02\pp ......\ple\, \nl
H$\beta$\tablenotemark{a}&
                  4861&  3.52\pp 0.28&       1.00\pp 0.00&       1.00\pp 0.00 \nl
[\ion{O}{3}]&     4959&  0.76\pp 0.12&       0.22\pp 0.04&       0.21\pp 0.04 \nl
[\ion{O}{3}]&     5007&  3.31\pp 0.16&       0.94\pp 0.09&       0.89\pp 0.08 \nl
\enddata
\tablenotetext{a}{Reddening-corrected flux is $F_0({\rm H}\beta) = (1.5\pm
		 0.2)\times 10^{-15}$ erg cm$^{-2}$ s$^{-1}$, and
		 luminosity is $L_0({\rm H}\beta) = (7.6\pm 1.0)\times
		 10^{38}$ erg s$^{-1}$ (for $H_0=75$).}
\label{tab4}
\end{deluxetable}
\clearpage


\begin{deluxetable}{lcccll}
\def\pnn{\phn\phn}
\def\psn{\phs\phn}
\def\pp{$\phantom{\pm}$}
\def\pla{$\phantom{\la}$}
\def\dts{....$\;$\pp ...$\:$}
\def\DTS{....$\;$\pp\pp ....$\phm{a}$}
\tablenum{5}				
\tablecolumns{6}
\tablewidth{0pt}
\tablecaption{Photometric vs Spectroscopic Cluster Ages}
\tablehead{
\colhead{} &   \colhead{$M_V$\tablenotemark{a}} &
\colhead{$(V-I)_0$\tablenotemark{b}} & 
\colhead{$\tau_{\rm phot}$} &   \colhead{} &   \colhead{$\:\tau_{\rm sp}$} \\
\colhead{Cluster} &   \colhead{(mag)} &   \colhead{(mag)} &
\colhead{(Myr)} &   \colhead{\pla$\log$($\tau_{\rm sp}/$yr)} &
\colhead{\ \,(Myr)} \\
\colhead{\ \,(1)} &   \colhead{(2)} &   \colhead{(3)} &
\colhead{(4)} &   \colhead{\pla(5)} &   \colhead{\ \,(6)}    }
\startdata
W3  & $-$16.2\,& $+0.64\pm 0.04$& $420\pm 110$     & \pla$8.73\pm 0.02$& $\phn 540\pm \phn30$  \nl
W6  & $-$14.4\,& $+0.64\pm 0.03$& $420\pm \phn{}90$& \pla$8.76\pm 0.04$& $\phn 580\pm \phn50$  \nl
W26 & $-$13.6\,& $+0.68\pm 0.03$& $510\pm \phn{}70$& \pla$8.72\pm 0.20$&
                                                              $\phn 530\,\,\phn_{-200}^{+300}$ \nl
W30 & $-$14.6\,& $+0.63\pm 0.03$& $400\pm \phn{}90$& \pla$8.67\pm 0.04$& $\phn 470\pm \phn40$  \nl
W31 & $-$13.0\,& $+0.54\pm 0.03$& $230\pm \phn{}40$& \pla$8.58\pm 0.23$&
                                                              $\phn 380\,\,\phn_{-160}^{+270}$ \nl
S101& $-$14.1: &    \nodata &       \nodata &                $\la$7.0  &          $\:\la$10    \nl
S105& $-$13.6: &    \nodata &       \nodata &        \pla$8.78\pm 0.08$\tablenotemark{c} &
                                                                             $\phn 600\pm 110$ \nl
S114& $-$12.8\,& $+0.62\pm 0.05$& $370\pm 110$ &     \pla$9.06\pm 0.10$\tablenotemark{d} &
                                                                         $1100\pm 300$     \nl
\enddata
\tablenotetext{a}{For $H_0 = 75$ km s$^{-1}$ Mpc$^{-1}$.}
\tablenotetext{b}{From Miller \etal\ (1997).}
\tablenotetext{c}{One of two possible values, other value being $8.36\pm 0.20$.}
\tablenotetext{d}{Uncertain; one of two possible values, other value being $7.6\pm 0.4$.}
\label{tab5}
\end{deluxetable}
\clearpage

%

\end{document}